\documentstyle[pra,aps,epsf]{revtex}

\begin{document}
\setcounter{table}{0}
\setcounter{figure}{0}
\twocolumn[\hsize\textwidth\columnwidth\hsize\csname@twocolumnfalse\endcsname

\title{Stability  of a
 vortex in a small trapped Bose-Einstein condensate}
\author{Marion Linn$^{1,2}$ and Alexander L.~Fetter$^1$}
\address{$^1$Department of Physics, Stanford
University, Stanford, CA 94305-4060 \\ $^2$Physikalisches
Institut, Universit\"at Bonn, Nu\ss allee 12, D-53115 Bonn,
Germany}

\date{\today}
\maketitle

\begin{abstract}

A second-order expansion of the Gross-Pitaevskii equation in the
interaction parameter determines  the thermodynamic
critical angular velocity  $\Omega_c$ for the creation of a vortex
in a small axisymmetric condensate. Similarly, a second-order
expansion of the Bogoliubov equations determines the (negative)
frequency $\omega_a$ of the anomalous mode. Although $\Omega_c =
-\omega_a$ through first order,   the second-order contributions
ensure that the absolute value $|\omega_a|$ is always smaller than
the critical angular velocity $\Omega_c$.  With increasing
external rotation $\Omega$, the dynamical instability  of the
condensate with a vortex disappears at $\Omega^*=|\omega_a|$,
 whereas the vortex state becomes energetically stable at the
larger value $\Omega_c$. Both  second-order  contributions depend
explicitly on the axial anisotropy of the trap. The appearance of a local
minimum  of the free energy for  a
vortex at the center determines the metastable angular velocity $\Omega_m$.  A
variational calculation yields
$\Omega_m=|\omega_a|$ to first order (hence $\Omega_m$ also coincides with
the critical angular velocity
$\Omega_c$ to this order). Qualitatively, the scenario for the
onset of stability  in the weak-coupling limit is the same as that
found in the strong-coupling (Thomas-Fermi) limit. \\


\end{abstract}
]

\tighten

\section{INTRODUCTION}

The experimental realization of Bose-Einstein condensates
\cite{And,Dav,Brad} in dilute alkali gases has stimulated great
interest in the study of vortices in these systems.  In the
absence of experimental detection of a condensate containing a
vortex, several authors have addressed the important question of
the stability of such a vortex
~\cite{Baym96,Dalf,Lundh,Dodd,Rokh,Fetter98,SF98,Svid98,Pu,IsoJ99,Garcia99,Feder99,Butts}.
The simplest criterion is energetic stability, determined by
comparing the total energy of a condensate with and without a
vortex, based on the time-independent Gross-Pitaevskii equation.
Separately, numerical and analytical studies of the Bogoliubov
equations for the excitations of a condensate with a vortex found
a normal mode with negative energy and positive normalization, suggesting a
dynamical instability~\cite{Dodd,Rokh,Fetter98,IsoJ99}. Moreover, in the
limit of large particle number, there is a metastable regime, for
which the free energy of the vortex condensate develops a local
minimum at the center of the trap, stabilizing the vortex against
small lateral displacements~\cite{Svid98,Feder99}. In the Thomas-Fermi
limit, the onset of metastability coincides with $|\omega_a|$,
 implying a close connection between the two
phenomena~\cite{Svid98}.

In the present work, we consider a small cylindrically symmetric
condensate with a central vortex line along the $z$ axis; this
weak-coupling limit allows a perturbation
expansion of the Gross-Pitaevskii (GP) equation. The additional
energy of the condensate with a vortex  can be compensated by
rotating the whole condensate with an angular velocity $\Omega_c$
\cite{Dalf}. The Bogoliubov equations for the same system have an
anomalous mode with negative frequency $\omega_a$.  To first order
in the (small) interaction parameter,
$|\omega_a|=\Omega_c$~\cite{Fetter98,Butts}. To understand the
possible connection between these two types of instabilities, we
here evaluate the second-order corrections to $\Omega_c$ and
$\omega_a$. Furthermore, the metastable frequency $\Omega_m$ computed for
the weak-coupling limit  agrees to first order with the
modulus of the anomalous mode, just as in the Thomas-Fermi limit.
This result indicates  that the instability with respect  to
microscopic oscillations and the onset metastability are closely related, for
 the system becomes dynamically stable when rotated faster than the
metastable frequency.

The following section uses the GP equation for a condensate with a
central vortex to determine the critical rotation frequency
$\Omega_c$ to second order. In the third section, a similar
perturbation expansion of the Bogoliubov equations for the
anomalous mode yields the second-order correction to the
corresponding eigenvalue of this mode $\omega_a$. The fourth section is
devoted to determining the metastable frequency $\Omega_m$
variationally, and  the conclusion discusses the implications of
our results.

\section{CRITICAL angular velocity}

Consider   a condensate containing $N$ particles in a harmonic
trap with radial and axial frequencies $\omega_\perp$ and
$\omega_z$. The interparticle interactions are characterized by a
positive  $s$-wave scattering length $a>0$.  The starting point
for the perturbation theory is the time-independent
Gross-Pitaevskii equation~\cite{Pit61,Gross61}
\begin{equation}
  (\,H_0 + 4 \pi \gamma\, |\Psi|^2\,)\Psi = \mu \,\Psi \, ,
\end{equation}
with $\mu$ the chemical potential.  Here,
\begin{equation}
H_0 =\case{1}{2} \bigg[-\nabla_\perp^2 + r^2+
\lambda\bigg(-{\partial^2\over
\partial z^2}+z^2\bigg)\bigg] \,
\end{equation}
is the Hamiltonian for the noninteracting condensate, expressed in
 dimensionless units (the radial and axial coordinates are
scaled with the radial and axial  oscillator lengths
$d_{\perp} = \sqrt{\hbar/m \omega_{\perp}}$ and $d_z =
\sqrt{\hbar/m \omega_z}$, and frequencies are scaled with the
radial trap frequency $\hbar\omega_\perp$). The condensate wave
function $\Psi$ is normalized to unity,    $\gamma \equiv  Na/d_z$
is the small interaction parameter  and
$\lambda=\omega_z/\omega_\perp$ represents the geometrical
asymmetry of the cylindrical trap.

The normalized eigenfunctions for the noninteracting condensate
 are taken as a product of the one-dimensional axial oscillator
state $\varphi_l(z)$ and the two-dimensional wave function for a
$q$-fold quantized vortex $\chi_{n+q,n}(r,\phi)$ at the center of the trap, so
that [compare Eqs.~(\ref{varphi}) and (\ref{chi}) in the appendix]
\begin{equation}
  \psi_{n+q,n,l} (\vec{r}) = \chi_{n+q,n}(r,\phi)\varphi_l(z) \, ,
\end{equation}
where $\psi_{q00}$ is the lowest energy state for a condensate
with a given quantized circulation $q$.  The noninteracting
single-particle states have an energy    $\epsilon_{n+q,n,l}^{(0)}
= 2n +q +1 +(l+\frac{1}{2})\lambda$.

The thermodynamic criterion for stability of a rotating condensate
with a $q$-fold vortex is the vanishing of  the free-energy
difference $\Delta F_q$ between the free energy of the condensate
with and without the vortex. The free energy for a $q$-fold vortex
state in the frame rotating with angular velocity $\Omega$ is $F_q
= E_q -\Omega q N$, where $E_q$ is the energy of the condensate
and $q$ is the angular-momentum quantum number. Setting $F_q-F_0
=0$ gives the critical rotation frequency

\begin{equation}
\Omega_c = \frac{E_q -E_0}{N q} \, .
\end{equation}

To construct the  perturbation theory,
the chemical potential and the condensate wave function are
expanded in the interaction parameter $\mu \approx
\mu^{(0)}+\gamma\mu^{(1)} + \cdots$ and $\Psi \approx
\Psi^{(0)}+\gamma\Psi^{(1)} + \cdots\,$.  The first-order correction
to the chemical potential  of the condensate is easily seen to be
$\mu^{(1)} = 4\pi\,\langle \Psi^{(0)} |\,|\Psi^{(0)}|^2
\,|\Psi^{(0)}
\rangle$, and the  thermodynamic relation
$\mu =
\partial E/\partial N$ then gives the first-order correction to the
condensate energy (note that $\gamma \propto N$)
\begin{equation}
  E^{(1)} = \case{1}{2} N\mu^{(1)} =2 \pi N \langle \Psi^{(0)}
|\,|\Psi^{(0)}|^2
\,|\Psi^{(0)}
\rangle \, .\label{eq:E1}
\end{equation}
For a $q$-fold vortex with  $\Psi^{(0)}=\psi_{q00}$,
direct evaluation of the matrix element gives
\begin{equation}
  \frac{E^{(1)}_q}{N} = \case{1}{2} \mu_q^{(1)}= \frac{(2q)!}{(q!)^2\,
2^{2q}\,\sqrt{2 \pi}\,} \,.\label{eq:mu1}
\end{equation}
The first-order correction to the energy  decreases with
increasing
$q$, and the corresponding critical rotation frequency

\begin{equation}
\Omega_c \approx 1-\frac{E_0^{(1)}-E_q^{(1)}}{Nq}\,\gamma
\end{equation}
increases
with increasing
$q$. Figure 1 shows the first-order approximation to the
thermodynamic critical angular velocity as a function of the
interaction parameter $\gamma$ for the first few values of
$q$. Since the resulting $\Omega_c$ is always smallest for $q=1$,
a singly quantized vortex will appear first, in which case $\Omega_c \approx
1-\frac{1}{2}\gamma/\sqrt{2\pi}$.   We consider only
 this case ($q=1$)  in determining the
second-order correction to the critical angular velocity.

\begin{figure}[hbt]\label{q}
\begin{center}
\epsfxsize=8cm\epsfbox{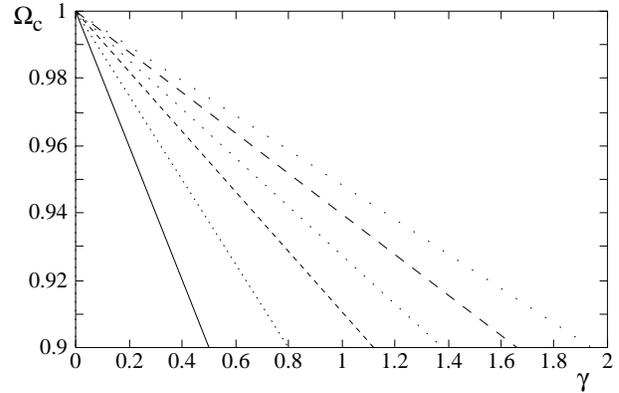} \vspace{10pt}
\caption{\narrowtext The critical rotation frequency in
first-order per\-tur\-bation theory for in\-crea\-sing quanta of
cir\-cu\-lation $q=1 \ldots 6$ (from left to right). The range of the
interaction parameter $\gamma$ is extended to unphysical values
for better illustration.}
\end{center}
\end{figure}

The second-order correction to the energy of the condensate can
be found by writing $\Psi^{(1)}$
 as a sum over the unperturbed
eigenfunctions of the appropriate symmetry (this form maintains
the normalization to first order)
\begin{equation}\Psi^{(1)}={\sum_{nl}}'c_{nl}\,
\chi_{n+q,n}\,\varphi_{2l}\,,\label{eq:Psi1a}
\end{equation}
where the prime on the sum indicates that it runs
over all nonnegative values of
$\{n,l\}$, omitting the single state $ \{0,0\}$.
Substitution into the GP equation and use of the orthogonality of
the  unperturbed solutions determine the coefficients
\begin{eqnarray}
c_{nl}&=& -\frac{2\pi}{n+\lambda l}\, \langle
\psi_{n+q,n,2l}|\,|\psi_{q00}|^2\,|\psi_{q00}\rangle \nonumber \\
&=&-\frac{2\pi}{n+\lambda l}\, I_{l}\,J^{nq}_{qqq}
\label{eq:Psi1b}
\end{eqnarray}
in terms of definite integrals discussed in the appendix. For
example, the axial integration over the product of four
harmonic-oscillator eigenfunctions yields
\begin{equation}I_l=\int_{-\infty}^\infty
dz\,\varphi_{2l}(z)\,\varphi_0(z)^3=
\frac{(-1)^l}{l!\,2^{2l}}\,\sqrt{\frac{(2l)!}{2\pi}}\,,\label{eq:Il}
\end{equation} with a similar but more complicated  expression
$J^{nq}_{ijk}$ for the   integral over the radial
eigenfunctions (they involve Laguerre polynomials).

 The corresponding second-order  correction to the condensate
energy follows immediately from the second-order chemical potential

\begin{eqnarray}
  \frac{E_q^{(2)}}{N} &=& \case{1}{3} \mu_q^{(2)} = 16\pi^2  \,{\sum_{nl}}'
  \,\frac{|\langle \psi_{n+q,n,2l}| \,|\psi_{q00}|^2 \,|\psi_{q00}
\rangle|^2}
  {\mu^{(0)}_q -\epsilon_{n+q,n,2l}^{(0)}} \nonumber \\
&=&
-8\pi^2\,{\sum_{nl}}'\,\frac{(I_{l}\,J^{n,q}_{qqq})^2}{n+\lambda
l}\, ,\label{eq:E2}
\end{eqnarray}
where $ \mu^{(0)}_q = \epsilon_{q00}^{(0)}= q + 1+
\frac{1}{2} \lambda$. The
angular quantum number $q$ is   one for the singly quantized vortex
or zero for the nonvortex ground state.

In contrast to the first-order correction $E_q^{(1)}$ in
Eq.~(\ref{eq:E1}), the second-order correction now depends
explicitly on the asymmetry parameter $\lambda$. The appendix
describes  a systematic way of writing the matrix elements
needed to carry out the infinite sums, and  the difference between
the second-order energy contributions for the vortex and nonvortex
state $\Delta E^{(2)}$ yields
 the second-order correction to the critical
rotation frequency

\begin{equation}
  \Omega^{(2)}_c(\lambda) =\frac{1}{16 \pi}\,{\sum_{nl}}'
  \frac{1}{n+\lambda
\,l}\,\frac{12 +3 n^2 -n^3}{\,2^{2n}}\,
  \frac{(2l)!}{2^{4l}\, (l!)^2} \, .
\end{equation}
Consequently the second-order approximation to the thermodynamic
critical rotation frequency (determined from the GP equation) is

\begin{equation}
  \Omega_c
  = 1 - \frac{1}{2\,\sqrt{2 \pi}}\,\gamma
  +\Omega^{(2)}_c(\lambda)\, \gamma^2 \,+\cdots .
\end{equation}

Table I lists  the second-order
contribution $\Omega_c^{(2)}(\lambda)$ for various asymmetry
parameters $\lambda$. This contribution is always positive, which
counteracts the (negative)   first-order contribution. For large
$\lambda$ (a disk-shaped condensate), $\Omega_c^{(2)}$ becomes
constant (from the terms with $l=0$), whereas for small $\lambda$
(a cigar-shaped condensate), it grows like $\Omega_c^{(2)} \propto
1/\lambda$ (from the terms with $n=0$). The resulting
thermodynamic critical angular velocity for several $\lambda$ is
shown in Fig.~2, illustrating the increasing importance of the
second-order term for small $\lambda$. For comparison, this figure
also includes the critical angular velocity for a $q=1$ vortex
with only the first-order correction taken into account.
\narrowtext
\begin{table}
\begin{tabular}{ddddd}
& $\lambda$   &  $\Omega_c^{(2)}$ & $-\omega_a^{(2)}$& \\
\tableline &$10^4$     &   0.0805  &   0.0508    & \\
 &$10$       &   0.0849  &   0.0546 &\\
 &$2 \sqrt{2}$&  0.0954  &   0.0637 &\\
 &$1$        &   0.1196  &   0.0851 &\\
 &$10^{-1}$  &   0.4227  &   0.3614 &\\
 &$10^{-2}$  &   3.4033  &   3.0933 &\\
 &$10^{-4}$  &   331.15  &   303.53 &
\end{tabular}
\vspace{10pt}
\caption{Second-order contributions for the critical rotation
frequency $\Omega_c^{(2)}$ and the anomalous Bogoliubov mode
$\omega_a^{(2)}$.} \label{table:comp}
\end{table}
\begin{figure}[hbt]\label{l}
\begin{center}
\epsfxsize=8cm\epsfbox{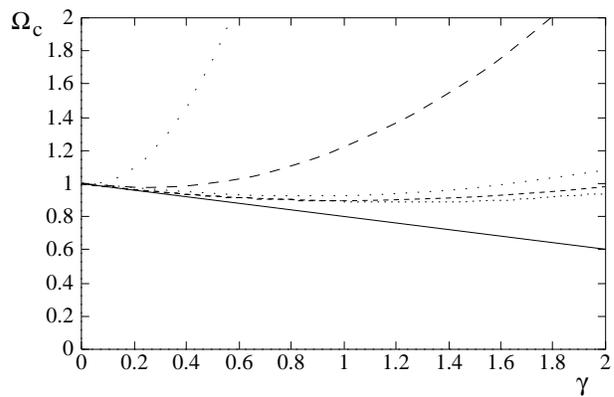} \vspace{10pt}
\caption{\narrowtext The critical rotation frequency in
second-order perturbation theory for decreasing asymmetry
parameters $\lambda=10, 2\sqrt{2},1,0.1,0.01$ (from bottom to
top). The solid line is $\Omega_c$ up to first order. Again,
$\gamma$ is extended beyond the range of validity for
illustration. }
\end{center}
\end{figure}

\section{ANOMALOUS MODE}

The preceding section dealt solely with the GP equation for the
condensate wave function $\Psi$ and the chemical potential $\mu$,
comparing the energy for a  condensate with and without  a vortex.
We now turn to the Bogoliubov equations~\cite{Bog47} that
describe the small-amplitude excitations of a condensate
containing a singly quantized vortex:

\begin{eqnarray}
\left(H_0 - \mu +8\pi \gamma\,|\Psi|^2\right)u_j -4 \pi \gamma\,
(\Psi)^2 v_j &=&  \omega_j u_j \, , \\
 -4 \pi \gamma\,
(\Psi^*)^2 u_j +\left(H_0 - \mu +8\pi \gamma\,|\Psi|^2\right)v_j
&=& - \omega_j v_j \, ,
\end{eqnarray}
where $u_j$ and $v_j$ are the normal-mode amplitudes and
$\omega_j$ is the corresponding frequency. In zero order, the
condensate wave function  $\Psi^{(0)}$ for a singly quantized
vortex is given by $\psi_{100} = \chi_{10}\varphi_0$ and the
associated chemical potential is $\mu^{(0)}=2
+\frac{1}{2}\lambda$. For physical solutions, the normal-mode
amplitudes have positive normalization $\int dV (|u_j|^2-|v_j|^2)
= 1$. Here, we focus on the anomalous mode, which  has a negative
eigenvalue $\omega_a <0$.  Hence  the creation of quasiparticles
in this mode lowers the energy relative to that of the condensate,
implying  a possible instability
~\cite{Dodd,Rokh,Fetter98,IsoJ99,Fetter99}.

 A noninteracting condensate with a singly quantized  vortex  is
unstable because the
 particles in the condensate with $\psi_{100}$ can make a
transition to the nonrotating condensate with $ \psi_{000}$,
giving up an energy $1$ in our dimensionless units. In the
presence of  interactions, however, the description
becomes more complicated, and the eigenfunctions of the Bogoliubov
equation include both $u_a$ and $v_a$, even in zeroth order.  At
this level, the Bogoliubov equations for the anomalous mode
become

\begin{eqnarray}
\left(H_0 - \mu^{(0)} \right)\,u_a^{(0)}  &=&
\omega_a^{(0)} u_a^{(0)} \, , \\
 \left(H_0 - \mu^{(0)} \right)\,v_a^{(0)}
&=& - \omega_a^{(0)} v_a^{(0)} \, ,
\end{eqnarray}
where the preceding argument suggests that $u_a^{(0)}\propto
\psi_{000}= \chi_{00}\varphi_0$ characterizes the nonrotating
vortex-free condensate,
 so that $\omega_a^{(0)} = -1$.  Since the full Bogoliubov
equation for
$u_a$ contains the coupling term $(\Psi)^2\,v_a\propto
e^{2i\phi}v_a$, it is natural to assume that $v_a\propto
e^{-2i\phi}$, and the unperturbed state $v_a^{(0)}\propto
\psi_{020}=\chi_{02}\varphi_0$ has the correct energy to satisfy
the remaining zero-order Bogoliubov equation.   To ensure the
proper normalization, we merely take
\begin{equation}
u_a^{(0)}= \cosh\theta\,\chi_{00} \varphi_0 \, \, \hbox{and}  \,
\, v_a^{(0)}= \sinh\theta\,\chi_{02}\varphi_0  \, ,\label{eq:zero}
\end{equation}
where the  parameter $\theta$ can only be determined by
including the higher-order terms in the perturbation expansion.

The first-order terms in the Bogoliubov equations become
\begin{eqnarray}
\left(H_0 - \mu^{(0)}\right) u_a^{(1)} &+&\left( 8\pi
\,|\Psi^{(0)}|^2-\mu^{(1)}\right)u_a^{(0)}  \nonumber \\ -4\pi\,
(\Psi^{(0)})^2 v_a^{(0)} &=&  \omega_a^{(0)}
u_a^{(1)}+\omega_a^{(1)} u_a^{(0)} \, ,\label{eq:u1}\\
\noalign{\vspace{.4cm}}
 \left(H_0 - \mu^{(0)}\right) v_a^{(1)} &+&\left( 8\pi
\,|\Psi^{(0)}|^2-\mu^{(1)}\right)v_a^{(0)}  \nonumber \\
-4\pi\, (\Psi^{(0)*})^2 u_a^{(0)} &=& - \omega_a^{(0)}
v_a^{(1)}-\omega_a^{(1)} v_a^{(0)}  \, ,\label{eq:v1}
\end{eqnarray}
and we must also expand the parameter {$\theta\approx
\theta^{(0)}+\gamma \theta^{(1)} + \cdots\,$}.  As a result, the
zero-order functions in Eq.~(\ref{eq:zero}) are evaluated with
$\theta^{(0)}$.
The first-order contributions then become
\begin{eqnarray}
u_a^{(1)} &=& \theta^{(1)}\sinh\theta^{(0)} \chi_{00}\,\varphi_0
 + {\sum_{nl}}' a_{nl} \,\chi_{nn}\,\varphi_{2l}\,,\label{eq:ua1}
\\
v_a^{(1)} &=& \theta^{(1)}\cosh\theta^{(0)}\chi_{02}\,\varphi_0 +
 {\sum_{nl}}'  b_{nl}\,
\chi_{n,n+2}\,\varphi_{2l}\, ;\label{eq:va1}
\end{eqnarray}
these expansions maintain the proper normalization to first order in
$\gamma$.

Multiply Eqs.~(\ref{eq:u1}) and (\ref{eq:v1}) by $u_a^{(0)*}$ and
$v_a^{(0)*}$, respectively, and integrate. Straightforward
manipulations yield the pair of equations
\begin{eqnarray}
\mu^{(1)}+\omega_a^{(1)}& =& 4\pi I_0\,\Big(2\int
d^2r\,|\chi_{00}|^2\,|\chi_{10}|^2  \nonumber \\
&&-\tanh\theta^{(0)}\int
d^2r\,\chi_{00}^*\,(\chi_{10})^2\,\chi_{02}\Big)\nonumber\\
&=&\frac{1}{\sqrt{2\pi}}\left(2-\frac{\tanh\theta^{(0)}}{\sqrt
2}\right)\,,\\ \mu^{(1)}-\omega_a^{(1)} &= &4\pi I_0\Big(2\int
d^2r\,|\chi_{02}|^2\,|\chi_{10}|^2 \nonumber \\ &&-
\coth\theta^{(0)}\int
d^2r\,\chi_{02}^*\,(\chi_{10}^{*})^2\,\chi_{00}\Big)\nonumber\\
&=&
\frac{1}{\sqrt{2\pi}}\left(\case{3}{2}-\frac{\coth\theta^{(0)}}
{\sqrt2}\right)\,,
\end{eqnarray}
where $I_0$ is given in Eq.~(\ref{eq:Il}), and the
 radial integrals are
 evaluated with the results in the appendix.
The sum of these equations determines $\mu^{(1)}$, and comparison
with Eq.~(\ref{eq:mu1}) yields
\begin{eqnarray}
\cosh\theta^{(0)}& = &\sqrt 2,\quad\hbox{and}\quad \nonumber \\
\sinh\theta^{(0)}&= &1.
\end{eqnarray}
Correspondingly, the difference now determines the first-order
correction to the anomalous frequency~\cite{Fetter98}
\begin{equation}
\omega_a^{(1)} = \frac{1}{2\sqrt{2\pi}}\,,
\end{equation}
so that $\Omega_c = -\omega_a$ through first order.

To determine the coefficients in the first-order expansions in
Eqs.~(\ref{eq:ua1}) and (\ref{eq:va1}), project Eqs.~(\ref{eq:u1})
and (\ref{eq:v1}) onto the appropriate unperturbed
eigenfunctions.  In this way, we find (see Appendix B)
\begin{eqnarray}
a_{nl}& =&\frac{2\pi\,I_l}{n+\lambda\, l} \,\Big(\int
d^2r\,\chi_{nn}^*\,(\chi_{10})^2\,\chi_{02}\nonumber \\ &&- 2\sqrt
2 \int
d^2r\,\chi_{nn}^*\,|\chi_{10}|^2\,\chi_{00}\,\Big)\nonumber\\
&=&\frac{2\pi\,I_l}{n+\lambda \, l}\left(J_{112}^{n0}-2\sqrt
2\,J_{110}^{n0}\right)\, ,\label{eq:anl}\\ b_{nl}&=&
\frac{2\pi\,I_l}{n+\lambda\, l} \,\Big(\sqrt 2\,\int
d^2r\,\chi_{n,n+2}^*\,|\chi_{10}|^2\,\chi_{02}\nonumber \\ &&- 2
\int
d^2r\,\chi_{n,n+2}^*\,(\chi_{10}^{*})^2\,\chi_{00}\,\Big)\nonumber\\
&=&\frac{2\pi\,I_l}{n+ \lambda \, l}\left(\sqrt 2
J_{112}^{n2}-2J_{110}^{n2}\right)\,.\label{eq:bnl}
\end{eqnarray}

The second-order
Bogoliubov equations are

\begin{eqnarray}
&&(\omega_a^{(0)}u_a^{(2)} + \omega_a^{(1)} u_a^{(1)} +
\omega_a^{(2)} u_a^{(0)}) \nonumber \\ \noalign{\vspace{.1cm}}&& =
(H_0- \mu^{(0)})\, u_a^{(2)} - \mu^{(1)} u_a^{(1)} - \mu^{(2)}
u_a^{(0)} \nonumber \\ && \, \,\, +8\pi\, [\,|\Psi^{(0)}|^2
u_a^{(1)} +
\left(\Psi^{(0)*}\Psi^{(1)}+\Psi^{(0)}\Psi^{(1)*}\right)u_a^{(0)}]
\nonumber
\\  && \, \, \, - 4 \pi \,[(\Psi^{(0)})^2 \,v_a^{(1)} + 2
\Psi^{(0)}\Psi^{(1)} v_a^{(0)}] \, , \label{bog1}\\
\noalign{\vspace{.4cm}} && - (\omega_a^{(0)}v_a^{(2)} +
\omega_a^{(1)} v_a^{(1)} + \omega_a^{(2)} v_a^{(0)})\nonumber \\
\noalign{\vspace{.1cm}}&&  = (H_0 -\mu^{(0)}) \,v_a^{(2)} -
\mu^{(1)} v_a^{(1)} - \mu^{(2)} v_a^{(0)} \nonumber
\\ &&\, \, \,+ 8 \pi \,[\,|\Psi^{(0)}|^2 v_a^{(1)} +
\left(\Psi^{(0)*}\Psi^{(1)}+\Psi^{(0)}\Psi^{(1)*}\right)v_a^{(0)}]
\nonumber \\
 && \, \, \,- 4 \pi \,[(\Psi^{(0)*})^2\,u_a^{(1)}
 + 2\Psi^{(0)*}\Psi^{(1)*}u_a^{(0)}] \, . \label{bog2}
\end{eqnarray}
Here, Eqs.~(\ref{eq:Psi1a}) and (\ref{eq:Psi1b}) give the
condensate wave function
$\Psi^{(1)}$, and Eqs.~(\ref{eq:ua1}), (\ref{eq:va1}),
(\ref{eq:anl}) and (\ref{eq:bnl}) give the corrections
$u_a^{(1)}$ and
$v_a^{(1)}$.

The remaining steps are essentially the same as in first order,
yielding a pair of equations for $\mu^{(2)}\pm\omega_a^{(2)}$.  The
parameter $\theta^{(1)}$ must be chosen so that the sum reproduces
 Eq.~(\ref{eq:E2}) for the chemical potential $\mu^{(2)}$.
The difference then gives  the  second-order correction to the
anomalous frequency
\begin{eqnarray} \omega^{(2)}_a(\lambda)
&=&8\pi^2{\sum_{nl}}' \frac{(I_l)^2}{n + l \lambda} \nonumber
\\ &\times&\Big[13(J_{111}^{n1})^2 - 8(J_{110}^{n0})^2
-4(J_{112}^{n2})^2 - 2(J_{110}^{n2})^2 \nonumber
\\ &&- (J_{112}^{n0})^2 - 8J_{100}^{n1}J_{111}^{n1}
-  4J_{122}^{n1}J_{111}^{n1}\nonumber \\ &&+ 4\sqrt{2}
(J_{112}^{n0}J_{110}^{n0} + J_{110}^{n2}J_{112}^{n2})\,\Big] \, ,
\end{eqnarray}
and  the total anomalous frequency through second order becomes
\begin{equation}
  \omega_a
 = -1 + {1\over 2}\frac{1}{\sqrt{2 \pi}}\,\gamma
  +\omega^{(2)}_a(\lambda) \,\gamma^2 \, .
\end{equation}
This is the main result of the present  section, illustrated in
Fig.~3.
\begin{figure}[hbt]\label{a}
\begin{center}
\epsfxsize=8cm\epsfbox{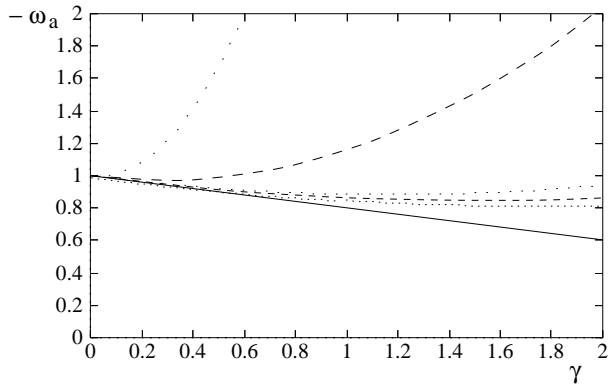}
\vspace{10pt}
\caption{\narrowtext The absolute values of the anomalous mode
frequency in second-order perturbation theory for decreasing
asymmetry parameters $\lambda=10, 2\sqrt{2},1,0.1,0.01$ (from
bottom to top). The solid line is $\omega_a$ up to first order.}
\end{center}
\end{figure}
 For several different
axial asymmetries,  Table I  compares the second-order correction
$\Omega_c^{(2)}(\lambda)$ to the critical rotation frequency from the
preceding section
 with the second-order correction $\omega_a^{(2)}(\lambda)$ to the
anomalous frequency.  It can be seen
that $\omega_a^{(2)}$ is always negative and smaller in absolute value
than the corresponding correction $\Omega_c^{(2)}$.

To understand the significance of this result,
consider a normal mode of the $q$-fold quantized vortex with
$u_j(r,\phi) = \exp[i(m_j +q)\phi]\, \tilde u_j(r)$ and $v_j(r,\phi)
=
\exp[i(m_j-q)\phi]\,\tilde v_j(r)$~\cite{Dodd,SF98}, where $m_j$
is the angular momentum of the normal mode
relative to that of the vortex in the condensate (namely, the corresponding
perturbations in the density and velocity potential have the angular dependence
$\propto\exp im_j\phi$). In a frame rotating with angular velocity
$\Omega$,   the frequency of a given normal mode
becomes  $\omega_j (\Omega) =
\omega_j (0) - m_j
\Omega$. For the present case of a singly quantized vortex with
$q=1$, the anomalous mode has
$m_a = -1$, so that the  shifted  anomalous
frequency in the rotating frame is
$\omega_a(\Omega) = \omega_a (0) +\Omega$. With increasing external
rotation $\Omega$, the condensate with a vortex becomes
dynamically stable against microscopic oscillations for
$\Omega\ge \Omega^* = |\omega_a|$.  Our numerical results show that
$\Omega^*<\Omega_c $,  so the onset of dynamical stability occurs
before
 the  singly quantized vortex in the condensate becomes energetically
favorable. Since the sum $\Omega_c+\omega_a$ vanishes through
order $\gamma$, it is necessary to include the second-order
corrections to decide  the relative value of $\Omega^*$ and
$\Omega_c$.
The difference is illustrated in Fig.~4 for two geometries,
including the corresponding results for strong coupling (in the
Thomas-Fermi limit)~\cite{Svid98}, which shows that the sequence of
the two types of stabilization is the same in both regimes.
\begin{figure}[hbt]\label{omega}
\begin{center}
\epsfxsize=8cm\epsfbox{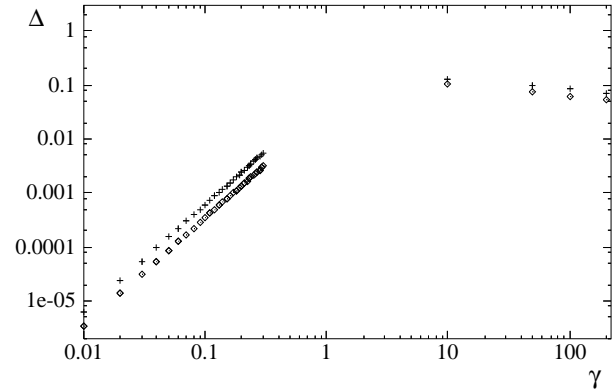} \vspace{10pt} \caption{\narrowtext
The difference $\Delta =\Omega_c-|\omega_a|$ (namely $\Omega_c
+\omega_a$) for the asymmetries $\lambda=1$(diamonds) and
$\lambda=0.1$(crosses). In the weak-coupling limit, the difference
is much smaller than in the Thomas-Fermi limit (note the
logarithmic scale), but remains positive.}
\end{center}
\end{figure}

\section{Metastable frequency}

The third important frequency in this problem is the metastable
angular velocity $\Omega_m$  at which a slightly off-center vortex becomes
trapped in a local minimum of the free energy at the center of the
confining potential. Consider the corresponding energy functional
\begin{equation}\label{efunc}
  E(\Psi) = \int dV \left[{\Psi^*}\left(H_0 -\Omega L_z\right)\Psi
  + 2 \pi \gamma |\Psi|^4 \right]\, \, .
\end{equation}
We now use cartesian coordinates, with
\begin{equation}\label{H0}
  H_0 = \frac{1}{2} \left[ - \partial^2_x- \partial^2_y + (x^2 +y^2)
  + \lambda \left( -\partial^2_z+z^2\right)\right]
\end{equation}
and $L_z = -i(x \partial y - y \partial x)$ is the $z$-component of
the angular-momentum operator.

In order to evaluate the energy functional Eq.~(\ref{efunc}), we
construct a trial condensate wave function. This trial function  is
assumed to be unchanged along the axis of symmetry, which allows us to
use the ground-state  gaussian $\varphi_0(z)$. In the weak-coupling
limit,  the radius of the vortex core is comparable with  the
radius of the condensate, so that the displacement ${\bf r}_0=(x_0,y_0)$
of the vortex, the displacement ${\bf r}_1=(x_1,y_1)$ of the condensate,
and the induced velocity of the condensate must all be included in the
following trial function
\begin{eqnarray}
  \psi_v(x,y,z)&=&\frac{C}{\pi^{3/4}}[(x-x_0)+i(y-y_0)] \nonumber \\
&\times& e^{-\case{1}{2}{(x-x_1)^2}-\case{1}{2}{(y-y_1)^2}-
\case{1}{2}{z^2}}
  e^{i(\alpha_x x+\alpha_y y)} \, ,
\end{eqnarray}
where $C^{-2} = 1+|{\bf r}_1-{\bf r}_0|^2$ and  the constants
$\alpha_x$ and $\alpha_y$ characterize the  velocity components.

In evaluating the integration in $E(\psi_v)$,
we retain all terms up to second order in the small parameters
$\alpha_i$ and in the displacements. For convenience, we introduce new
 variables $\bbox{\delta}={\bf
r}_1-{\bf r}_0 =(\delta_x,\delta_y)$ and $\bbox{\epsilon}=2{\bf
r}_1-{\bf r}_0 =(\epsilon_x,\epsilon_y)$, which will turn out to
be the normal modes of the system.  The variational  energy is

\begin{eqnarray}\label{Ecross}
E_{\rm var} (\bbox{\alpha},\bbox{\delta},\bbox{\epsilon} \,)
&=& 2+{\lambda \over 2} -\Omega +{\gamma\over
2\sqrt{2\pi}}\nonumber \\
 &+&{\alpha^2 \over 2} +\alpha_x\,(-\delta_y+\Omega\epsilon_y)
+\alpha_y\,(\delta_x-\Omega\epsilon_x) \nonumber \\
&+&{\epsilon^2 \over 2} -\Omega \, \bbox{\delta}\cdot\bbox{\epsilon}
+\Big(2\Omega-{3\over2}+{\gamma\over \sqrt{2\pi}}\Big)\delta^2.
\end{eqnarray}
The values
\begin{eqnarray}
 \alpha_x &=& \delta_y -\Omega \epsilon_y \\
  \alpha_y &=& -\delta_x +\Omega \epsilon_x
\end{eqnarray}
minimize this expression  with respect to the velocity parameter
$\bbox{\alpha} =(\alpha_x,\alpha_y)$,
and  the  energy then becomes diagonal in
the  variables $\bbox{\delta}$ and $\bbox{\epsilon}$, which thus represent the
appropriate  normal modes
\begin{eqnarray}\label{Ede}
 E_{\rm var}(\bbox{\delta},\bbox{\epsilon} \,) &=& 2 +\frac{\lambda}{2}+
\frac{\gamma}{2\sqrt{2 \pi}} - \Omega  \nonumber \\
 &+&\left( -2 +\frac{\gamma}{\sqrt{2 \pi}} + 2 \Omega \right){\delta}^2
 +\frac{\left(1- \Omega^2\right)}{2}{\epsilon}^2  \, .
\end{eqnarray}

 The energy of the
vortex-free ground state is $E_0 = 1+\lambda /2+ \gamma /(\sqrt{2
\pi})$.  Thus  the difference $\Delta F(0) = E_{\rm var}(0) -E_0$ vanishes at
the expected critical angular velocity  $\Omega_c = 1 -\gamma/(2 \sqrt{2
\pi})$, in agreement with the first-order result obtained from the
Gross-Pitaevskii equation.  Furthermore, if $\bbox{\delta}$ vanishes, then the
vortex and condensate move rigidly, and the resulting dipole oscillation mode
is stable throughout the range $|\Omega|\le 1$.  For larger angular velocity,
the motion becomes unstable, as is familiar from the behavior of a
classical particle in a parabolic potential.

We next consider how  the relative displacement
$\bbox{\delta}$ affects the variational energy  Eq.~(\ref{Ede}),
which is clearly unstable if
$\Omega=0$.  With increasing external rotation, however, the relative
displacement becomes metastable at a frequency
\begin{equation}\label{Om}
\Omega_m = 1 -\frac{\gamma}{2\sqrt{2 \pi}} \, ,
\end{equation}
 when $E_{\rm var}$ changes from a local maximum to a local minimum for small
relative displacements.
To this (first) order in the interaction parameter $\gamma$, the value
$\Omega_m$  coincides with the modulus of the anomalous mode frequency
$|\omega_a|$ and hence with the critical frequency
$\Omega_c$.

In order to illustrate the importance of the combined effect  of the
vortex and the condensate,  we can study the behavior
if  only the vortex is  displaced (namely ${\bf r}_1=0$)
or if the induced velocity is neglected ($\bbox{\alpha}=0$). In the former
case, the energy (\ref{Ede}) is given by
\begin{eqnarray}\label{Ex0}
 E_{\rm var}({\bf r}_0) &=& 2 +\frac{\lambda}{2}+ \frac{\gamma}{2\sqrt{2 \pi}}
-
\Omega  \nonumber \\
 &+&\left( -\frac{3}{2} +\frac{\gamma}{\sqrt{2 \pi}} + 2 \Omega
-\frac{\Omega^2}{2}\right)
 \left(x_0^2 +y_0^2\right),
\end{eqnarray}
and the metastable frequency is determined by requiring a positive
coefficient of the displacement contribution [the second line in
Eq.~(\ref{Ex0})]. To first order in the interaction parameter, we
find \begin{equation}
\Omega_m^* = 1 -\frac{\gamma}{\sqrt{2 \pi}},\label{Omegam*}
\end{equation}
which differs from the value $\Omega_m$ found with the more general approach.
In the second case, the metastable frequency is determined by the condition
that the last line in Eq.~(\ref{Ecross}) is positive, namely  the determinant
of the coefficients must be positive. This gives the same metastable frequency
(\ref{Omegam*}) as that found from Eq.~(\ref{Ex0}). Therefore, the metastable
frequency has the same value as the
absolute value of the anomalous mode (up to first order) only if the
displacement of the condensate \emph{and} the induced velocity are both taken
into account. This indicates that
the vortex becomes confined in a local central energy minimum at the same
rotation  frequency $|\omega_a|$ for which the instability due to the
anomalous mode disappears. The same scenario had been found in the
strong-coupling limit~\cite{Svid98}, and our results strengthen the idea
of a common underlying phenomenon.

Note that the metastable frequency found here [Eq.~(\ref{Om})]
does not coincide with the metastable rotation frequency defined by Feder,
Clark, and Schneider~\cite{Feder99}. Instead of requiring a local minimum in
the free energy, they identify the onset of metastability  with the
frequency for which the chemical potentials for a condensate with
and without a vortex are equal. In first-order perturbation
theory, their criterion leads to the expression (\ref{Omegam*})
found by omitting the displacement of the condensate.

\section{CONCLUSIONS}

It has been shown that the critical rotation frequency $\Omega_c$
exceeds the modulus of the anomalous Bogoliubov mode $\omega_a$
when second-order corrections in the interaction parameter are included. This
result  agrees with numerical results for one particular trap geometry found
by Feder, Clark, and Schneider~\cite{Feder99}.
Furthermore, the sequence of stabilizing a singly quantized vortex
through rotation is found to be the same as in the strong-coupling
limit~\cite{Svid98}: First, the vortex becomes stable against
microscopic oscillations at $\Omega^* = |\omega_a|$, and only at
the higher rotation speed $\Omega_c$ does the vortex become
energetically stable. The geometry dependence introduced through
the second-order terms indicates that pancake geometries are more
favorable for vortex detection in rotating traps, since the
stabilization frequencies are lower. This is consistent with the
numerical results of  Garc\'{\i}a-Ripoll and
P\'erez-Garc\'{\i}a~\cite{Garcia99}. For extreme cigar-shaped condensates, the
critical frequency can even exceed the trap frequencies~\cite{Feder99}, a
regime that is experimentally inaccessible.  The angular velocity $\Omega_m$
for the onset of metastability coincides with the modulus of the anomalous
frequency $\omega_a$ in first order, when the combined effect of the vortex
and the condensate is taken into account. For $\Omega >\Omega_m$, an energy
barrier stabilizes the vortex at the center; at the same frequency, the
instability due to microscopic oscillations disappears. These features in the
stability scenario also agree with the behavior in the Thomas-Fermi limit
\cite{Svid98}.

\acknowledgements

We thank A.~A.~Svidzinsky for va\-lua\-ble
discussions. This work was supported in part by NSF Grant
No.~94-21888 and by the DAAD (German Academic Exchange Service)
``Doktorandenstipendium im Rahmen des gemeinsamen
Hochschulsonderprogramms III von Bund und L\"andern" (M.~L.).

\appendix

\section{Noninteracting eigenstates}

The eigenstates for the noninteracting Bose condensate in a
cylindrical trap can be classified with the quantum numbers of
positive and negative circulation $n_+,n_-$ around the $z$ axis
 and the axial harmonic-oscillator energy quantum number
$l$ in the
$z$ direction~\cite{Cohen}.  In particular, the $z$-dependent parts
of the eigenfunctions  are simple
harmonic oscillator eigenfunctions

\begin{equation}\label{varphi}
\varphi_l(z) =\frac{1}{\sqrt{ \pi^{1/2}\,2^l
l!}}\,H_l(z)
\,e^{-\frac{1}{2} z^2} \, ,
\end{equation}
where the $H_l(z)$ are the Hermite polynomials~\cite{Lebedev}.

In terms of the circular quanta, the
normalized two-dimensional eigenfunctions  are

\begin{equation}
\chi_{n_+,n_-}(x,y) =
\frac{1}{\sqrt{\pi\,n_+!\,n_-!}}\left(a_+^\dagger\right)^{n_+}
\left(a_-^\dagger\right)^{n_-}e^{-\frac{1}{2}(x^2 +y^2)}  \, ,
\end{equation}
where $a_\pm^\dagger= (a_x^\dagger \pm ia_y^\dagger )/\sqrt2=
\frac{1}{2}[x\pm iy-(\partial_x\pm i\partial_y)]$ are the creation
operators for right and left circular quanta, respectively.
In terms of the  new variables
$\zeta = x+i y$, $\zeta^* = x-i y $,
these operators take the form

\begin{equation}
a_+^\dagger = \frac{\zeta}{2} - \frac{\partial}{\partial\zeta^*}\,
,\qquad a_-^\dagger =
\frac{\zeta^*}{2} - \frac{\partial}{\partial\zeta} \, .
\end{equation}
The identity
$(\frac{1}{2}\zeta^*-\partial_\zeta)\exp(-\frac{1}{2}\zeta\zeta^*)
=\exp(\frac{1}{2}\zeta\zeta^*)
(-\partial_\zeta)$ $\exp(-\zeta\zeta^*)$ and its complex conjugate
readily yield
\begin{eqnarray}
\chi_{n+q,n} &=&
\frac{e^{\case{1}{2}\zeta\zeta^*}}{\sqrt{\pi\,(n+q)!\,n!}}
\left(-\frac{\partial}{\partial\zeta}\right)^n\,
\left(-\frac{\partial}{\partial\zeta^*}\right)^{n+q}\,
e^{-\zeta\zeta^*}\nonumber\\
&=&\frac{(-1)^n \,e^{\case{1}{2}\zeta\zeta^*}}{\sqrt{\pi\,(n+q)!\,n!}}
\left(\frac{\partial}{\partial\zeta}\right)^n\,
\left(\zeta^{n+q}e^{-\zeta\zeta^*}\right)\,.
\end{eqnarray}
Comparison with the standard formula for the associated Laguerre
polynomials $L_n^q$~\cite{Lebedev}
 yields

\begin{equation}\label{chi}
\chi_{n+q,n}(r,\phi) =
(-1)^{n}\sqrt{\frac{n!}{\pi\,(n+q)!}}\,e^{-\frac{1}{2}r^2}
\,e^{iq\phi}\,r^q
\,L_n^q(r^2) \, .
\end{equation}
The complete normalized three-dimensional eigenfunctions of the
noninteracting system are
$\psi_{n+q,n,l} (\vec{r}) = \chi_{n+q,n}(r,\phi)\varphi_l(z)$.

\section{Second-order matrix elements}

 The above expressions allow us to evaluate  the necessary  matrix
elements, and the  $z$-dependent part factorizes  out  in the form
\begin{eqnarray}
I_l &=& \int_{-\infty}^\infty dz\,\varphi_{2l}(z)\,\varphi_0(z)^3 \nonumber \\
&=&
\frac{1}{\pi\sqrt{2^{2l}\,(2l)!}}\int_{-\infty}^\infty dz\,H_{2l}(z)
e^{-2z^2},
\end{eqnarray}
where only states with an
even number of quanta contribute.  The generating function for the
Hermite polynomials
$$e^{-t^2+2zt} = \sum_{n=0}^\infty\frac{H_n(z)}{n!}\,t^n$$
readily shows that $\exp(-\frac{1}{2}t^2)$
is the generating function for $\sqrt{2^{2l}\,(2l)!}\,I_l$, and a
straightforward analysis gives the desired expression

\begin{equation}\label{B}
I_l =
\frac{1}{\sqrt{2 \pi}} \frac{(-1)^l \sqrt{(2 l)!}}{2 ^{2l} l!} \,.
\end{equation}

For the radial part note that $\chi_{n,n+q}(r,\phi) =
\chi_{n+q,n}^*(r,\phi)$. Since the phase factor is the only
complex part, it can be separated explicitly

\begin{eqnarray}
 \chi_{n+q,n}(r,\phi) &=&e^{iq\phi} \tilde{\chi}_{n+q,n}(r) \, ,\nonumber \\
 \chi_{n,n+q}(r,\phi) &=& e^{-iq\phi}\tilde{\chi}_{n+q,n}(r) \, ,
\end{eqnarray}
where $\tilde\chi $ is real.
The matrix elements $ J^{nq}_{ijk}=\int d^2r\,
\tilde{\chi}_{n+q,n}\,\tilde{\chi}_{i,0}\,\tilde{\chi}_{j,0}\,
\tilde{\chi}_{k,0}$ involve a product of four of
these eigenfunctions, where three refer to the condensate in the
lowest energy state, namely $n=0$ (note that $L_0^\alpha=1$ for any
$\alpha$).  Due to the angular integration, only products with no net
overall phase remain. Use of Eq.~(\ref{chi})  yields

\begin{equation}
J^{nq}_{ijk}= \frac{(-1)^n}{ \pi} \,\frac{1}{\sqrt{i!\,j!\,k!}}
\,
\sqrt{\frac{n!}{ (n+q)!}}\,\int_0^\infty  du \,u^{p} e^{-2u}
L^q_n(u)
\, ,
\end{equation}
where $p = \frac{1}{2}(q+i+j+k)$ is an integer because of the angular phase
factors (note also that
$J_{ijk}^{nq}$ is symmetric under interchange of its subscripts).
As in the preceding
example of the axial matrix elements, the generating function  for
the Laguerre polynomials~\cite{Lebedev}
$$\sum_{n=0}^{\infty} L_n^q (u) \,t^n =
\frac{1}{(1-t)^{1+q}}\, \exp\left(\frac{-u t}{1-t}\right)$$
facilitates the
radial integration.
 For example the function $2(1-t)/(2-t)^3$ provides a
generating function for $(-1)^n\,\pi\,\sqrt{n+1}\,J_{111}^{n1}$,
which is the matrix element with four singly quantized vortex
eigenfunctions, and we find

\begin{equation}\label{J}
  J^{n1}_{111} =
 \frac{(-1)^n}{2^{n+3}\,\pi}\,\sqrt{n+1}\,(2-n) \, .
\end{equation}
A similar technique leads to all the other relevant radial
integrals.

The sums occurring in the second-order terms contain
 the square of the matrix elements $(J^{nq}_{ijk}
\,I_l)^2$ appropriately weighted with the energy denominator. For
example, the second-order contribution to the chemical potential
for the singly quantized vortex  (obtained from the
Gross-Pitaevskii equation) is

\begin{eqnarray}\label{mu2}
  \mu_1^{(2)} &=&-24\pi^2\, {\sum_{nl}}'
  \frac{ (J^{n1}_{111}\, I_l)^2}{n + l\lambda}
   \nonumber \\
  &=&-\frac{3}{16 \pi}\,{\sum_{nl}}'
  \frac{1}{n+\lambda l}\frac{(n+1)(2-n)^2}{2^{2n}}
  \frac{(2l)!}{2^{4l} (l!)^2}\, .
\end{eqnarray}

\end{document}